\documentclass[copyright,creativecommons]{eptcs}
\providecommand{\event}{CompMod 2009} 
\providecommand{\volume}{??}
\providecommand{\anno}{2009}
\providecommand{\firstpage}{1}
\providecommand{\eid}{??}

\usepackage{verbatim}
\usepackage{graphicx}
\usepackage{amsmath, amssymb}
\usepackage{pepa}
\usepackage{txfonts}
\usepackage{empheq}
\usepackage{setspace}
\usepackage{array}

\newcommand{\Real}{{\mathbb R}}

\newcommand{\reactant}{\mbox{$\downarrow$}}
\newcommand{\product}{\mbox{$\uparrow$}}

\newcommand{\modifier}{\odot}
\newcommand{\activator}{\oplus}
\newcommand{\inhibitor}{\ominus}

\title{Quantifying the implicit process flow abstraction in SBGN-PD diagrams with Bio-PEPA}
\author{Laurence Loewe$^{1}$ \qquad\qquad Stuart Moodie$^{2,1}$ \qquad\qquad Jane Hillston$^{2,1}$
\institute{$^{1}$Centre for System Biology at Edinburgh\\
$^{2}$School of Informatics\\
The University of Edinburgh\\ Scotland}
\email{Laurence.Loewe@ed.ac.uk \qquad     Stuart.Moodie@ed.ac.uk \quad Jane.Hillston@ed.ac.uk    }
}
\def\titlerunning{SBGN-PD Process Flow Abstraction}
\def\authorrunning{Loewe, Moodie \& Hillston}

\begin{document}

\maketitle

\begin{abstract}
For a long time biologists have used visual representations of biochemical networks to gain a quick overview of important structural properties. Recently SBGN, the Systems Biology Graphical Notation, has been developed to standardise the way in which such graphical maps are drawn in order to facilitate the exchange of information. Its qualitative Process Diagrams (SBGN-PD) are based on an implicit Process Flow Abstraction (PFA) that can also be used to construct quantitative representations, which can be used for automated analyses of the system. Here we explicitly describe the PFA that underpins SBGN-PD and define attributes for SBGN-PD glyphs that make it possible to capture the quantitative details of a biochemical reaction network. 
We implemented SBGNtext2BioPEPA, a tool that demonstrates how such quantitative details can be used to automatically generate working Bio-PEPA code from a textual representation of SBGN-PD that we developed. Bio-PEPA is a process algebra that was designed for implementing quantitative models of concurrent biochemical reaction systems. We use this approach to compute the expected delay between input and output using deterministic and stochastic simulations of the MAPK signal transduction cascade. The scheme developed here is general and can be easily adapted to other output formalisms.
\end{abstract}

\section{Introduction}
\label{Sec:Introduction}

To describe biological pathway information visually has many advantages and  Systems Biology Graphical Notation Process Diagrams (SBGN-PD \cite{SBGN}) have been developed for this purpose. Like electronic circuit diagrams, they aim to unambiguously describe the structure of a complex network of interactions using graphical symbols. 
To achieve this requires both a collection of symbols and rules for their valid combination. SBGN-PD is a visual language with a precise grammar that builds on an underlying abstraction as the basis of its semantics (see p.40 \cite{SBGN}). We call this underlying abstraction for SBGN-PD the ``Process Flow Abstraction'' (PFA). It describes biological pathways in terms of processes that transform elements of the pathway from one form into another. The usefulness of an SBGN-PD description critically depends on the faithfulness  of the underlying PFA and a tight link between the PFA and the glyphs used in diagrams.
The graphical nature of SBGN-PD allows only for qualitative descriptions of biological pathways. However, the underlying PFA is more powerful and also forms the basis for quantitative descriptions that could be used for analysis. Such descriptions, however, need to allow the inclusion of the corresponding mathematical details like parameters and equations for computing the frequency with which reactions occur.

Here we aim to make explicit the PFA that already underlies SBGN-PD implicitly. This serves a twofold purpose. First, a better and more intuitive understanding of the underlying abstraction will make it easier for biologists to construct SBGN-PD diagrams. Second, the PFA is easily quantified and making it explicit can facilitate the quantitative description of SBGN-PD diagrams. Such descriptions can then be used directly for predicting quantitative properties of the system in simulations. Here we demonstrate how this could work by mapping SBGN-PD to a quantitative analysis system. We use the process algebra Bio-PEPA \cite{CiocHill09,BioPEPA} as an example, but our mapping can be easily applied to other formalisms as well. 

The rest of the paper is structured as follows. First we provide an overview of the implicit PFA with the help of an analogy to a system of water tanks, pipes and pumps (Section 2). In Section 3 we explain how this system can be extended in order to capture quantitative details of the PFA. We then show how SBGN-PD glyphs can be mapped to a quantitative analysis framework, using the Bio-PEPA modelling environment \cite{BioPEPA} as an example (Section 4). In Section 5 we discuss various internal mechanisms and data structures needed for translation into any  quantitative analysis framework.
As an example of how the translation process works, we apply our new translation tool ``SBGNtext2BioPEPA''  \cite{SBGNtextWebsite,SBGNtextTechReport} to a simple model of the MAPK signalling cascade \cite{Huang96}, which we automatically translate into Bio-PEPA, where we analyse the stochastic behaviour of the time needed for the cascade to be switched from ``off'' to ``on''. We end by reviewing related work and providing some perspectives for further developments.

\begin{small}
\begin{figure}
\label{figPFA}
\begin{center}
\includegraphics[width=1.0\textwidth] {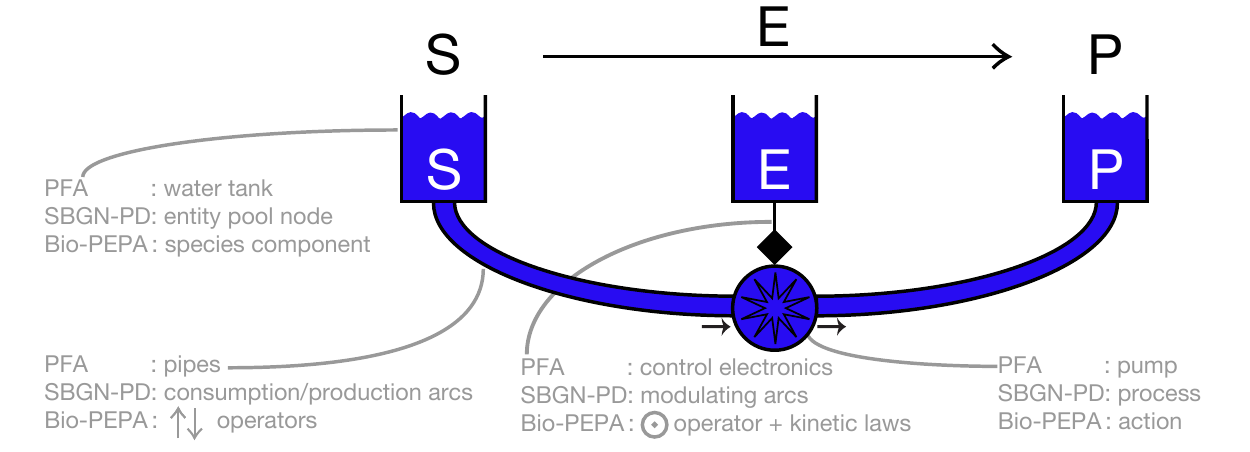}
\caption{An overview of the process flow abstraction. The chemical reaction at the top is translated into an analogy of water tanks, pipes and pumps that can be used to visualise the process flow abstraction. The various elements are also mapped into SBGN-PD and Bio-PEPA terminology. } 
\end{center}
\end{figure}
\end{small}

\section{The implicit Process Flow Abstraction of SBGN-PD}
\label{Sec:PFA}
The PFA behind SBGN-PD is best introduced in terms of an analogy to a system of many water tanks that are connected by pipes. Each pipe either leads to or comes from a pump whose activity is regulated by dedicated electronics. In the analogy, the water is moved between the various tanks by the pumps. In a biochemical reaction system, this corresponds to the biomass that is transformed from one chemical species into another by chemical reactions. SBGN-PD aims to also allow for descriptions at levels above individual chemical reactions. Therefore the water tanks or chemical species are termed ``entities'' and the pumps or chemical reactions are termed ``processes''. For an overview, see Figure 1.
We now discuss the correlations between the various elements in the analogy and in SBGN-PD in more detail. In this discussion we occasionally allude to SBGNtext, which is a full textual representation of the semantics of SBGN-PD (developed to facilitate automated translation of SBGN-PD into other formalisms; see \cite{SBGNtextWebsite,SBGNtextTechReport}). Here are the key elements of the PFA:

\begin{description}
  \item \textbf{Water tanks = entity pool nodes (EPNs).} Each water tank stands for a different pool of entities, where the amount of water in a tank represents the biomass that is bound in all entities of that particular type that exist in the system. Typical examples for such pools of identical entities are chemical species like metabolites or proteins. SBGN-PD does not distinguish individual molecules within pools of entities, as long as they are within the same compartment and identical in all other important properties.   An overview of all types of EPNs (i.e. categories of water tanks) in SBGN-PD is given in Table 1. To unambiguously identify an entity pool in SBGNtext and in the code produced for quantitative analysis, each entity pool is given a unique \texttt{EntityPoolNodeID}. The PFA does not conceptually distinguish between non-composed entities and entities that are complexes of other entities. Despite potentially huge differences in complexity they are all ``water tanks'' and further quantitative treatment does not treat them differently.

\begin{small}
\begin{table}
\label{reactionsEPNs}
\centering  
\extrarowheight3pt
\begin{tabular}{c  c  c  c}
\hline
SBGN-PD glyph &  \verb"EPNType"  & class type  & comment   \\[3pt]
\hline 

 \begin{minipage}[c] {1.8cm}   
\includegraphics[width=0.7\textwidth] {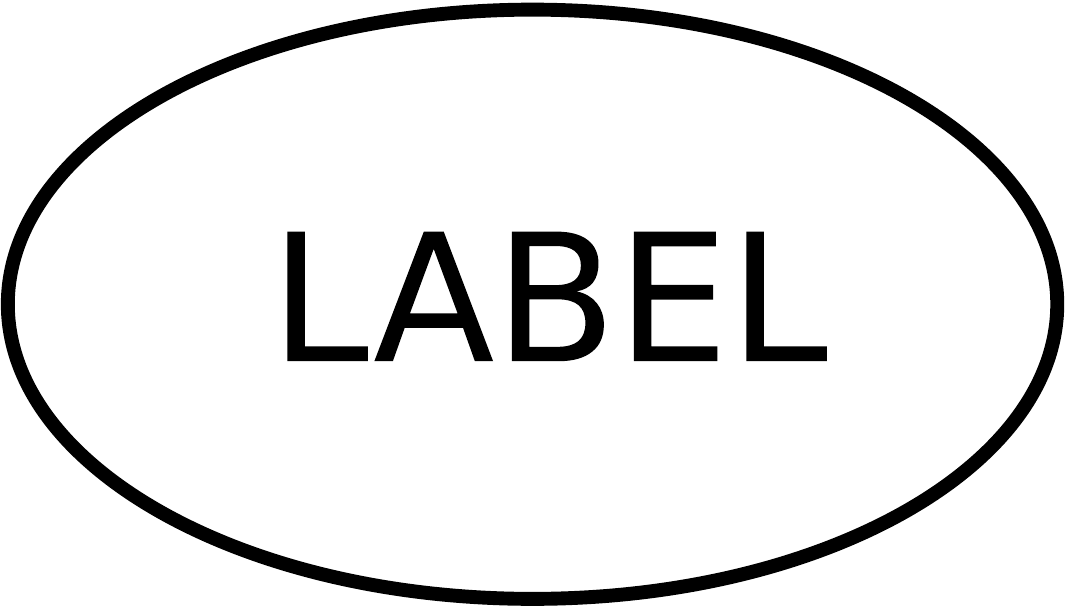}
\end{minipage}  &
\verb"Unspecified" &  material &   EPN with unknown specifics  \\

 \begin{minipage}[c] {1.8cm} 
\includegraphics[width=0.3\textwidth] {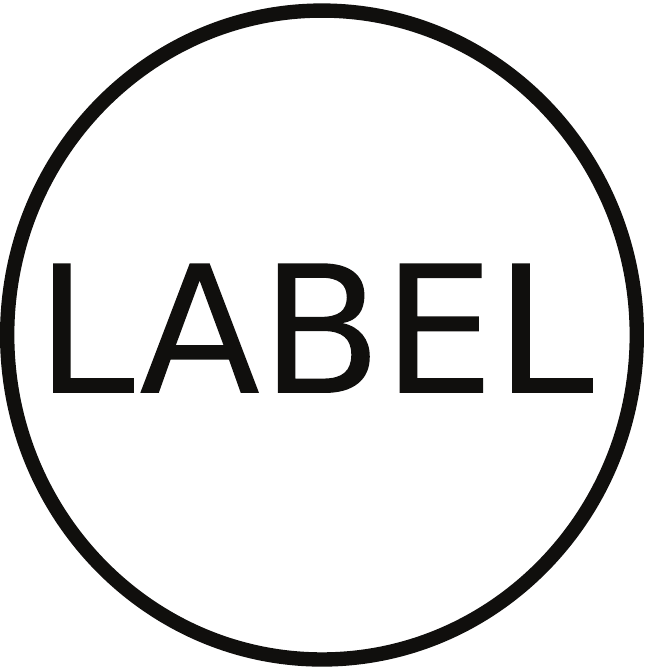}
\end{minipage}  &
\verb"SimpleChemical" &  material &    EPN   \\

 \begin{minipage}[c] {1.8cm} 
\includegraphics[width=0.7\textwidth] {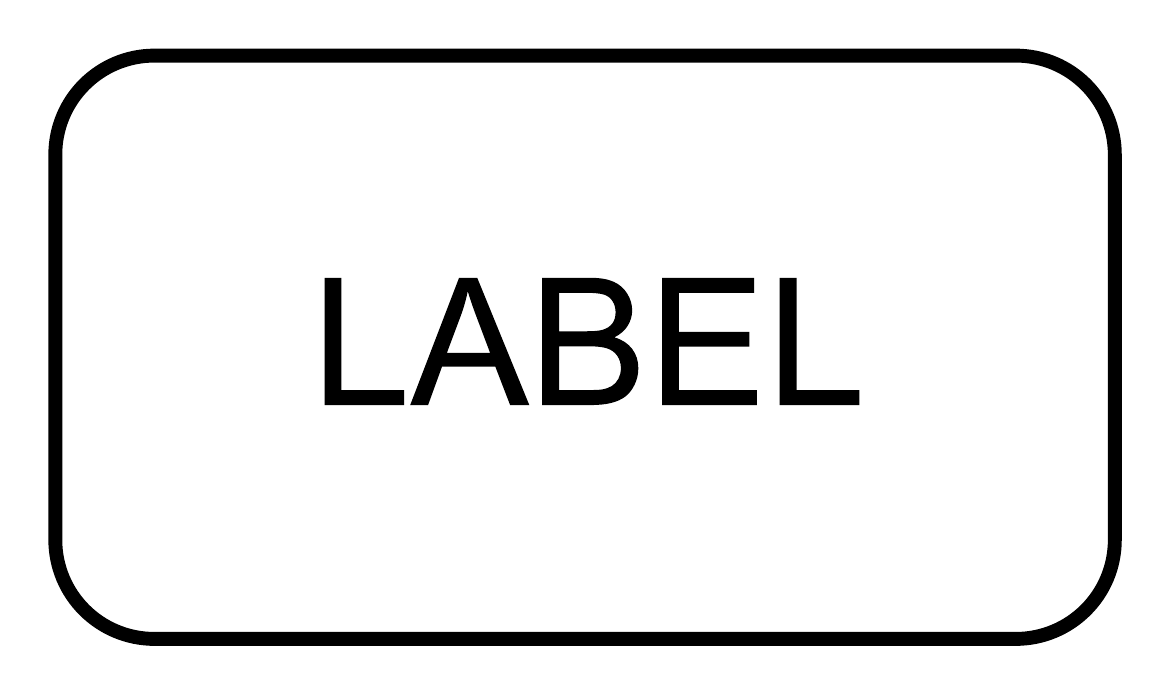}
\end{minipage}  &
\verb"Macromolecule" &  material &   EPN   \\

 \begin{minipage}[c] {1.8cm} 
 \includegraphics[width=0.7\textwidth] {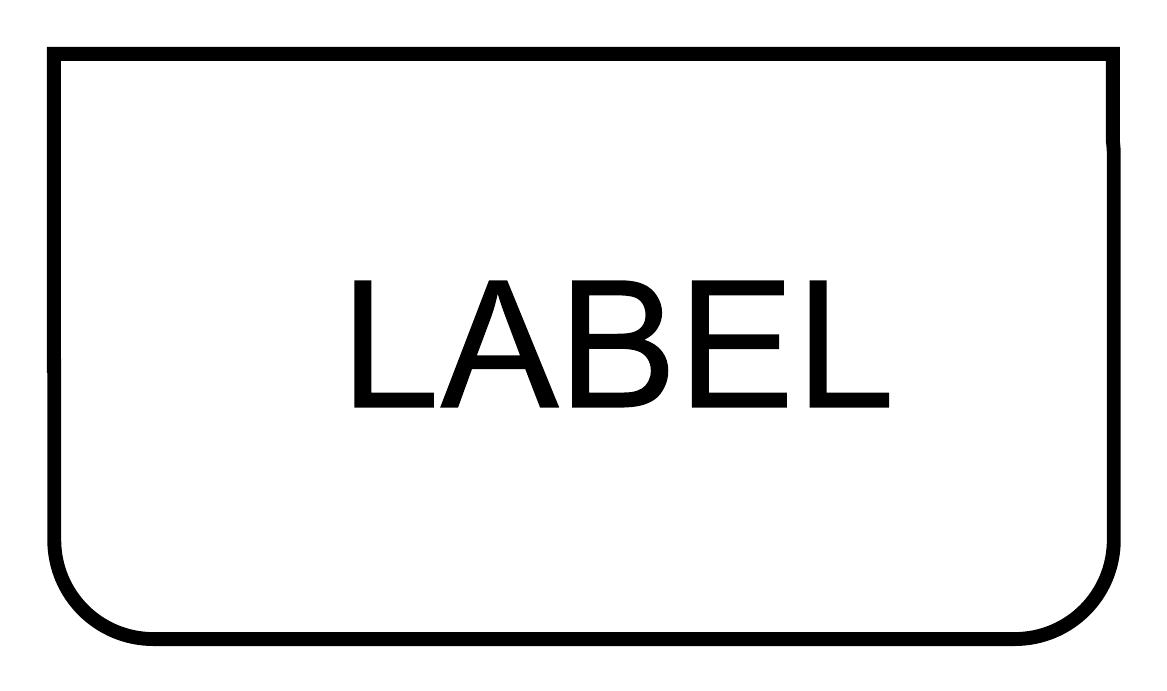}
\end{minipage}  &
\verb"NucleicAcidFeature" &  material &    EPN   \\

 \begin{minipage}[c] {1.8cm} 
\includegraphics[width=1.0\textwidth] {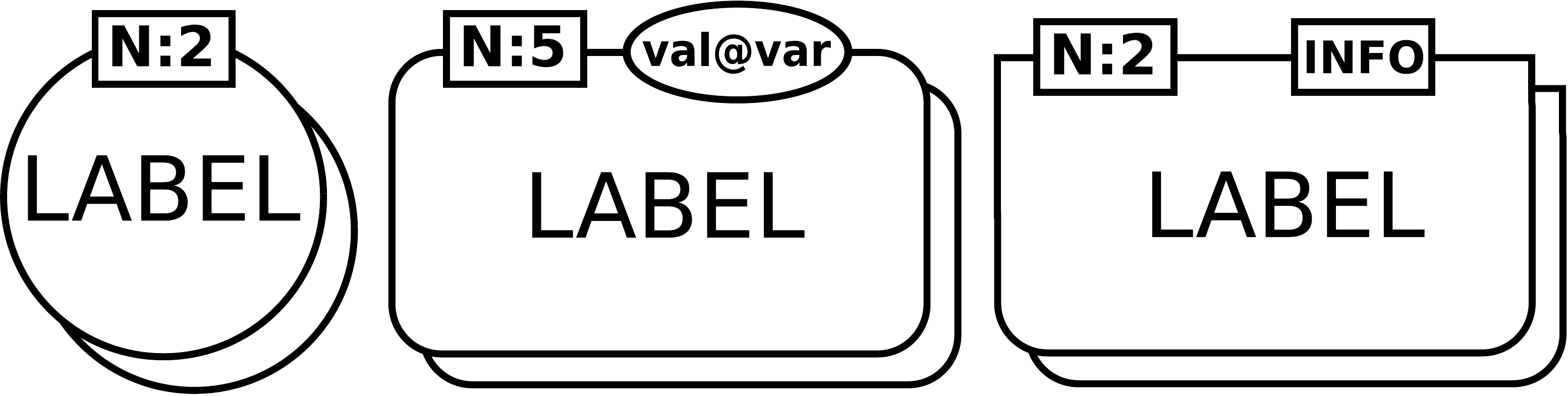}
\end{minipage}  & - 
 &  material &   EPN multimer specified by cardinality  \\[5pt]

 \begin{minipage}[c] {1.8cm} 
\includegraphics[width=1.0\textwidth] {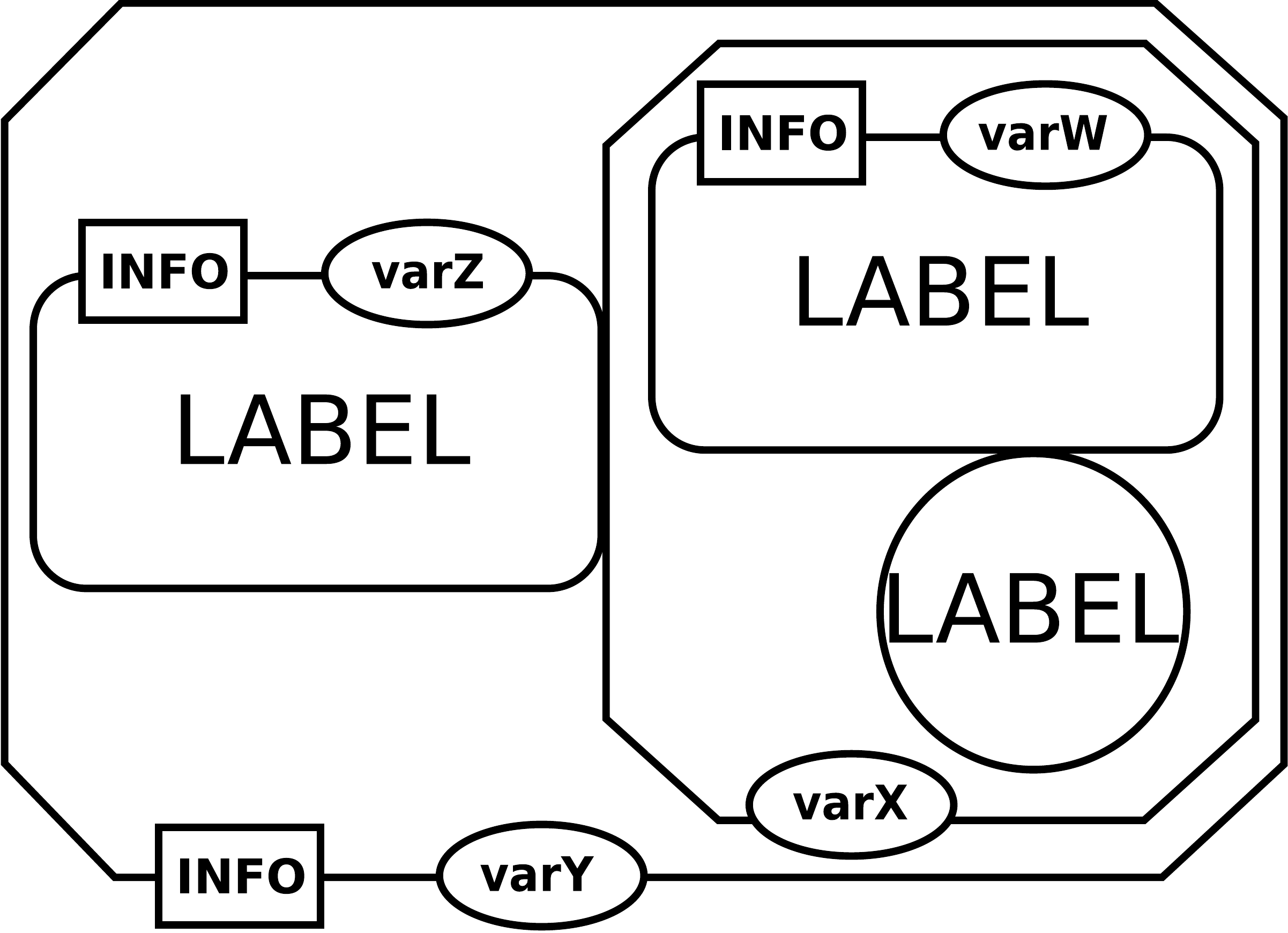}
\end{minipage}  &
\verb"Complex" &  container &  EPN;  arbitrary nesting allowed  \\

 \begin{minipage}[c] {1.8cm} 
\includegraphics[width=0.25\textwidth] {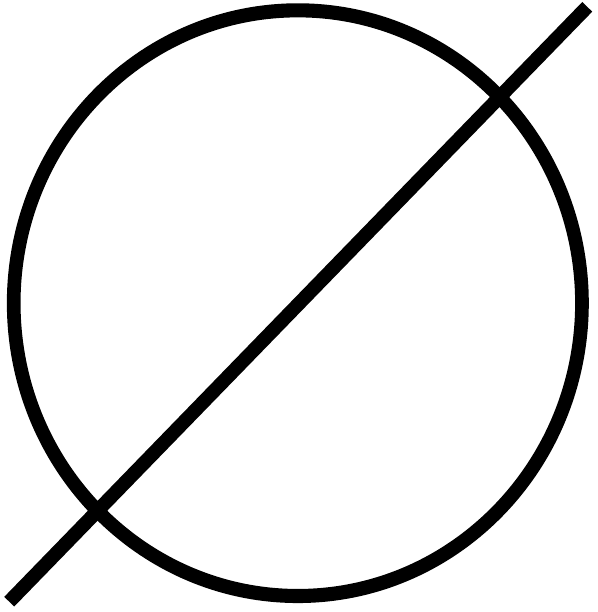}
\end{minipage}  &
\verb"Source" &  conceptual &  external source of molecules  \\

 \begin{minipage}[c] {1.8cm} 
\includegraphics[width=0.25\textwidth] {Table-Glyph-sourceSink.pdf}
\end{minipage}  &
\verb"Sink" &  conceptual &  removal from the system  \\

 \begin{minipage}[c] {1.8cm} 
\includegraphics[width=0.7\textwidth] {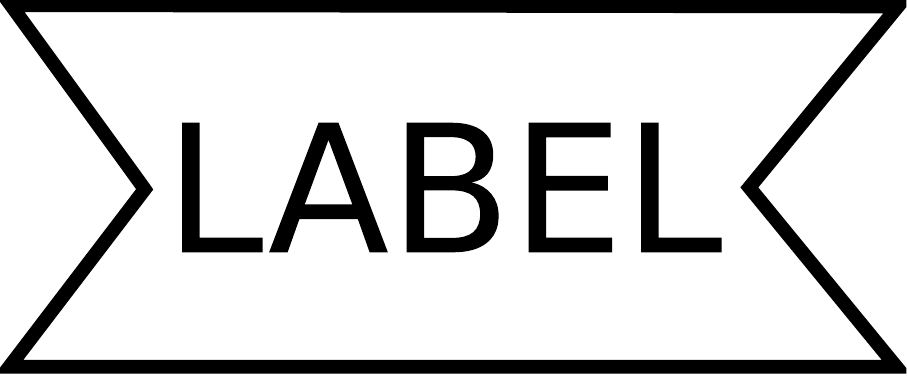}
\end{minipage}  &
\verb"PerturbingAgent" &  conceptual &  external influence on a reaction  \\[5pt]
\hline & &  \\[-7pt]

 \end{tabular}
\caption{ Categories of ``water tanks'' in the PFA correspond to types of entity pool nodes in SBGN-PD. The complex and the multimers are shown with exemplary auxiliary units that specify cardinality, potential chemical modifications and other information.} 
\end{table}
\end{small}

  \item \textbf{Pipes = consumption and production arcs.} Pipes allow the transfer of water from one tank to another. Similarly, to move biomass from one entity pool to another requires the consumption and production of entities as symbolised by the corresponding arcs in SBGN-PD (see Table 3, page \pageref{ArcsMapToBioPEPA}). These arcs connect exactly one process and one EPN.  The thickness of the pipes could be taken to reflect  stoichiometry, which is the only explicit quantitative property that is an integral part of SBGN-PD.  Production arcs take on a special role in reversible processes by allowing for bidirectional flow.
    
\item \textbf{Pumps = processes.} Pumps move water through the pipes from one tank to another. Similarly, processes transform biomass bound in one entity to biomass bound in another entity, i.e., processes transform one entity into another. The speed of the pump in the analogy corresponds to the frequency with which the reaction occurs and determines the amount of water (or biomass) that is transported between tanks (or that is converted from one entity to another, respectively).
Processes can belong to different types in SBGN-PD (Table 2) and are unambiguously identified by a unique \texttt{ProcessNodeID} in SBGNtext. This allows arcs to clearly define which process they belong to and by finding all its arcs, each process can also identify all EPNs it is connected to.

 {\em Reversible processes.} 
SBGN-PD allows for processes to be reversible if they are symmetrically modulated  (p.28 \cite{SBGN}).  
 Thus, there may be flows in two directions, however the net flow at any given time will be unidirectional. The PFA does not prescribe how to implement this. For simplicity, our analogy assumes pumps to be unidirectional, like many real-world pumps. Thus bidirectional processes in our analogy are represented as two pumps with corresponding sets of pipes and opposite directions of flow. 
In our implementation we follow SBGN-PD in separating  the left hand side and right hand side of reversible processes for an unambiguous description of reality if all relevant arc glyphs look like production arcs  (p.32 \cite{SBGN}). For more details, see \cite{SBGNtextTechReport}.

\begin{small}
\begin{table}
\label{reactions}
\centering  
\extrarowheight3pt
\begin{tabular}{c  c   c}
\hline
SBGN-PD glyph & \;\;\;\; \verb"ProcessType"  & \;\;\; meaning   \\[3pt]
\hline 

 \begin{minipage}[c] {1.8cm}   
\includegraphics[width=1\textwidth] {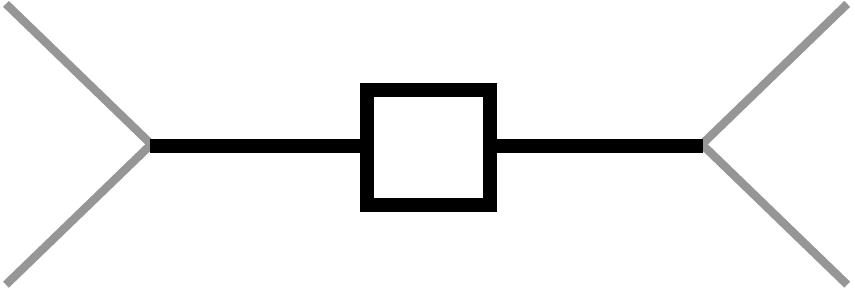}
\end{minipage}  &\;\;\;\;
\verb"Process" &  \;\;\;\; normal known processes  \\[2pt]

 \begin{minipage}[c] {1.8cm} 
 \includegraphics[width=0.8\textwidth] {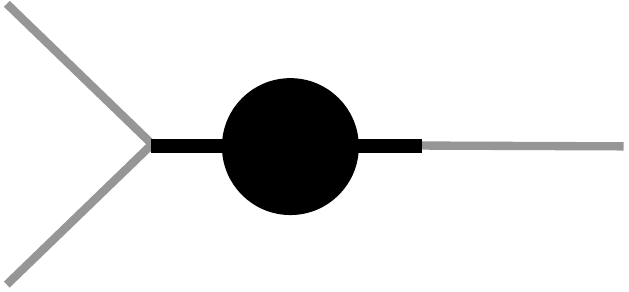}
\end{minipage}  &\;\;\;\;
\verb"Association" &  \;\;\;\; special process  that builds complexes \\

 \begin{minipage}[c] {1.8cm} 
 \includegraphics[width=0.8\textwidth] {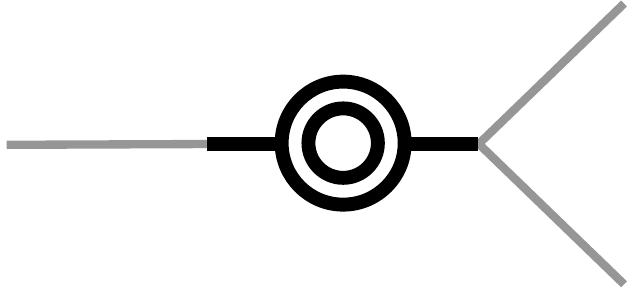}
\end{minipage}  &\;\;\;\;
\verb"Dissociation" &  \;\;\;\;special process  that dissolves complexes   \\

 \begin{minipage}[c] {1.8cm} 
\includegraphics[width=1\textwidth] {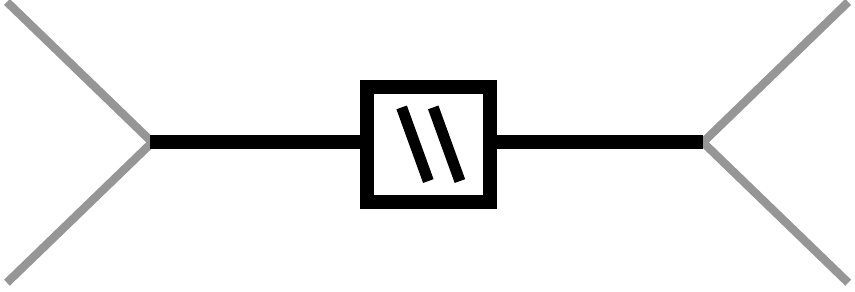}
\end{minipage}  &\;\;\;\;
\verb"Omitted" & \;\;\;\; several known processes are abstracted    \\[2pt]

 \begin{minipage}[c] {1.8cm} 
\includegraphics[width=1\textwidth] {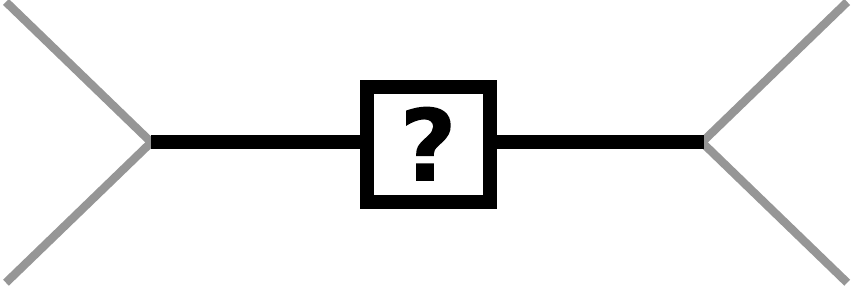}
\end{minipage}  &\;\;\;\;
\verb"Uncertain" &  \;\;\;\;existence of this process is not clear   \\ [-3pt]

 \begin{minipage}[c] {1.8cm} 
\includegraphics[width=1\textwidth] {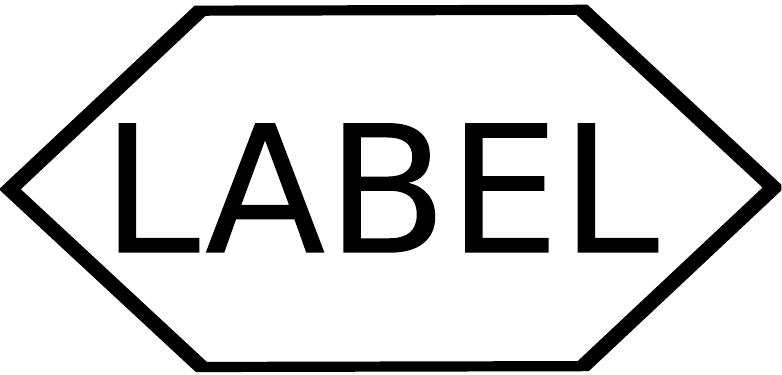}
\end{minipage}  &\;\;\;\;
\verb"Observable" &  \;\;\;\;this process is easily observable   \\[7pt]
\hline & &  \\[-7pt]

 \end{tabular}
\caption{ Categories of ``pumps'' in the PFA correspond to types of processes in SBGN-PD.} 
\end{table}
\end{small}

\item \textbf{Control electronics for pumps = modulating arcs and logic gates.} In the analogy, pumps need to be regulated, especially in complex settings. This is achieved by control electronics. In SBGN-PD, the same is done by various types of modulation arcs, logic arcs and logic gates \cite{SBGN}. They all contribute to determining the frequency of the reaction. Since SBGN-PD does not quantify these interactions, most of our extensions for quantifying SBGN-PD address this aspect. 
Each arc connects a ``water tank'' with a given \texttt{EntityPoolNodeID} and a ``pump'' with a given \texttt{ProcessNodeID}. Ordinary modulating arcs can be of type \verb"Modulation" (most generic influence on reaction), \verb"Stimulation" (catalysis or positive allosteric regulation), \verb"Catalysis" (special case of stimulation, where activation energy is lowered), \verb"Inhibition" (competitive or allosteric) or \verb"NecessaryStimulation" (process is only possible, if the stimulation is ``active'', i.e. has surpassed some threshold). The glyphs are shown in Table 3 (page \pageref{ArcsMapToBioPEPA}), where their mapping to Bio-PEPA is discussed. 
One might misread SBGN-PD to suggest that Consumption/Production arcs cannot modulate the frequency of a process. However, kinetic laws frequently depend on the concentration of reactants, implying that these arcs can also contribute to the ``control electronics'' (e.g. report ``level of water in tank'').
Another part of the ``control electronics'' are \emph{logical operators}. These simplify modelling, when a  biological function can be approximated by a simple on/off logic that can be represented by boolean operators. SBGN-PD supports this simplification by providing the logical operators ``AND'', ``OR'' and ``NOT'', which are connected by ``logic arcs'' with the rest of the diagram (logic arcs convert to and from the non-boolean world). 

    \newpage
    
    \item \textbf{Groups of water tanks = compartments, submaps and more.} The PFA is complete with all the elements presented above. However, to make SBGN-PD more useful for modelling in a biological context, SBGN-PD has several features that make it easier for biologists to recognise various subsets of entities that are related to each other. For example, entities that belong to the same compartment can be grouped together in the compartment glyph and functionally related entities can be placed on the same submap.  In the analogy, this corresponds to grouping related water tanks together. SBGN-PD also supports sophisticated ways for highlighting the inner similarities between entities based on a knowledge of their chemical structure (e.g. modification of a residue, formation of a complex).  Stretching the analogy, this corresponds to a way of highlighting some similarities between different water tanks. None of these groupings are important for the PFA in principle or for quantitative analysis, as long as different ``water tanks'' remain separate.
    
\end{description}

\section{Extensions for quantitative analysis}
\label{Sec:Extensions}

The process flow abstraction that is implicit in all SBGN process diagrams can be used as a basis to quantify the systems they describe. After we made the PFA explicit above, we now discuss the attributes that need to be added to the various SBGN-PD glyphs in order to allow for automatic translation of SBGN-PD diagrams into quantitative models. 
These attributes  are stored as strings in SBGNtext (our textual representation of SBGN-PD, see \cite{SBGNtextTechReport}) and are attached to the corresponding glyphs by a graphical SBGN-PD editor. They do not require a visual representation that compromises the visual ease-of-use that SBGN-PD aims for. 
Next we discuss the various attributes that are necessary for the glyphs of SBGN-PD to support quantitative analysis. We do not discuss auxiliary units, submaps, tags and equivalence arcs here, as they do not require extensions for supporting quantitative analysis.

\subsection{Quantitative extensions of EntityPoolNodes}
\label{sec:ExtEntityPoolNode}

For quantitative analysis, each unique EPN requires an \verb"InitialMoleculeCount" to unambiguously define how many entities exist in this pool in the starting state. We followed developments in the SBML standard in using counts of molecules instead of concentrations, since SBGN-PD also allows for multiple compartments, which makes the use of concentrations very cumbersome (see section 4.13.6, p.71f.  in \cite{SBML}).  
 For entities of type \verb"Perturbation", the \verb"InitialMoleculeCount" is interpreted as the numerical value associated with the perturbation, even though its technical meaning is not a count of molecules. 
Entities of the type \verb"Source" or \verb"Sink" are both assumed to be effectively infinite, so \verb"InitialMoleculeCount" does not have a meaning for these entities. 
Beyond a unique \texttt{EntityPoolNodeID} and  \texttt{Initial\-Molecule\-Count}, no other information on entities is required for quantitative analysis.

\subsection{Quantitative extensions of Arcs}
\label{sec:extend-arc}

 Arcs link entities and processes by storing their respective IDs and the \verb"ArcType". The simplest arcs are of type \verb"Consumption" or \verb"Production" and do not require numerical information beyond the stoichiometry that is already defined in SBGN-PD as a property of arcs that can be displayed visually in standard SBGN-PD editors. Logic arcs will be discussed below. 
All modulating arcs are part of the ``control electronics'' and affect the frequency with which a process happens. They link to EPNs to inform the process about the presence of enzymes, for example. Modulation is usually governed by parameters or other important quantities for the given process (e.g. Michaelis-Menten-constant). 

To make the practical encoding of a model easier, we define process parameters that conceptually belong to a particular modulating entity as a list of \verb"QuantitativeProperties" in the arc pointing to that entity. This is equivalent to seeing the set of parameters of a reaction as something that is specific to the interaction between a particular modulator and the process it modulates. Other approaches are also possible, but lead to less elegant implementations. Storing parameters in equations requires frequent and possibly error-prone changes (e.g. many different Michaelis-Menten equations). One could also argue that the catalytic features are a property of the enzyme and thus make parameters part of EPNs; however this either forces all  Michaelis-Menten reactions of an enzyme to happen at the same speed or requires cumbersome naming conventions to manage different affinities for different substrates.

To refer to parameters we specify the \verb"ManualEquationArcID" of an arc and then the name of the parameter that is stored in the list of \verb"QuantitativeProperties" of that arc.
This scheme reduces clutter by limiting the scope of the relevant namespace (only few arcs per process exist, so \verb"ManualEquationArcID"s only need to be unique within that immediate neighbourhood). Thus parameter names can be brief, since they only need to be unique within the arc.  
The \verb"ManualEquationArcID" is specified by the user in the visual SBGN-PD editor and differs from ArcID, a globally unique identifier that is automatically generated by the graphical editor. 
The \verb"ManualEquationArcID"  allows for user-defined generic names that are easy to remember, such as 'Km' and 'vm' for Michaelis-Menten reactions. It should be easily accessible within the graphical editor, just as the parameters that are stored within an arc.

\emph{ Logical operators and logic arcs.}
To facilitate the use of logical operators in quantitative analyses one needs to convert the integer molecule counts of the involved EPNs to binary signals amenable to boolean logic. 
Thus SBGNtext supports ``incoming logic arcs'' that connect a ``source entity'' or 
``source logical result'' with a ``destination logic operator'' and apply an ``input threshold'' to decide whether the source is above the threshold (``On'') or below the threshold (``Off'').  To this end, a graphical editor needs to support the ``input threshold'' as a numerical attribute that the user can enter; all other information recorded in incoming logic arcs is already part of an SBGN diagram. 
Once all signals are boolean, they can be processed by one or several logical operators, until the result of this operation is given in the form of either 0 (``Off'') or 1 (``On'').  This result then needs to be converted back to an integer or float value that can be further processed to compute process frequencies. Thus a graphical editor needs to support corresponding attributes  for defining a low and a high output level.

\subsection{Quantitative extensions of ProcessNodes}
\label{sec:Ext ProcessNodes}
For quantitative analyses, a ProcessNode must  have a unique name and an equation that computes the propensity, which is proportional to the probability that this process occurs next, based on the current global state of the model. 
 Since the \verb"ProcessType" is not required for quantitative analyses, it does not matter whether a process is an ordinary \verb"Process", an \verb"Uncertain" process or an \verb"Observable" process, for example.  For all these ProcessNodes, graphical editors need to support attributes for the manual specification of a \texttt{ProcessNodeID},  and a \verb"PropensityFunction". These attributes are then stored in SBGNtext. If support for bidirectional processes is desired, then graphical editors need to  facilitate entering a propensity function for the backward process as well. 
Propensity functions compute the propensity of a unidirectional process to be the next event in the model and can be used directly by simulation algorithms and solvers \cite{Gillespie07}. 

To instantiate the propensity function, a translator needs to replace all aliases by their true identity. We use the following syntax for a parameter alias that is substituted by  the actual numeric value (or a globally defined parameter) from the corresponding arc:

$$ \verb"<par: ManualEquationArcID.QuantitativePropertyName> "$$
While translating to Bio-PEPA this would be simply substituted with a corresponding parameter name. The parameter is then defined elsewhere in the Bio-PEPA code to have  the numerical value stored in the corresponding property of the arc. 
To allow the numerical analysis tool to access an EPN count at runtime we replace the following entity alias by the \texttt{EntityPoolNodeID} that the corresponding arc links to:
$$ \verb"<ent: ManualEquationArcID > "$$
This is shorter than the \texttt{EntityPoolNodeID} and allows the reuse of propensity functions if kinetic laws are identical and the manual IDs follow the same pattern.
It is desirable that there is no need to specify the \texttt{EntityPoolNodeID}.  It is fairly long and generated automatically to reflect various properties that make it unique. It would be cumbersome to refer to in the equation and it would require a mechanism to access the automatically generated \texttt{EntityPoolNodeID} before a SBGNtext file is generated. Also any changes to an entity that would affect its \texttt{EntityPoolNodeID} would then also require a change in all corresponding propensity functions, a potentially error-prone process. The same substitution mechanism can be used to provide access to properties of compartments (see \cite{SBGNtextTechReport}). 

 In addition to these aliases, functions use the typical standard arithmetic rules and operators that are merely handed through to the analysis tool.

\section{Mapping SBGN-PD elements to Bio-PEPA}
\label{Sec: Mapping SBGN-PD to Bio-PEPA}

In this section we explain how to use the semantics of SBGN-PD to map a SBGN-PD model to a formalism for the quantitative analysis of biochemical systems. We are using Bio-PEPA as an example, but our approach is general and can be applied to many other formalisms.

\subsection{The Bio-PEPA language} 
\label{}

 Bio-PEPA is a stochastic process algebra which models biochemical
pathways as interactions of distinct entities representing reactions of chemical species \cite{CiocHill09,BioPEPA}.  
A process algebra model captures the behaviour of a system as the
actions and interactions between a number of entities, where the latter are often termed ``processes'', ``agents'' or ``components''. In PEPA and Bio-PEPA these are built up from simple 
\emph{sequential components} \cite{CiocHill09,Hillston96,CalderHill09}.  Different process algebras support different modelling styles for biochemical systems \cite{CalderHill09}. Stochastic process
algebras, such as PEPA \cite{Hillston96} or the stochastic
$\pi$-calculus \cite{Priami95}, associate a random variable with each action to represent the mean of its exponentially distributed 
waiting time.   In the stochastic $\pi$-calculus, interactions are strictly binary whereas in Bio-PEPA the more general multiway synchronisation is supported.
The syntax of Bio-PEPA is defined as \cite{CiocHill09} :
 
 $$
S ::= (\alpha, \kappa) \mbox{ \texttt{op} } S \mid S + S \mid C
\quad \quad \quad P::=P \sync{\mathcal{L}}   P \mid S(x)   
$$

\noindent  where $S$ is a sequential \emph{species component} that represents a chemical
species (termed ``process'' in some other process algebras and ``EntityPoolNode'' in SBGN-PD), $C$ is a constant pointing to an  $S$, $P$ is a \emph{model component} that
describes  the set $\mathcal{L}$ of possible interactions between species  components (these ``interactions'' or ``actions'' correspond to ``processes'' in SBGN-PD and can represent chemical reactions).  A  count of molecules or a concentration of $S$ is given by $x \in \Real^+_0$.
In the prefix term ``$(\alpha,\kappa) \mbox{ \texttt{op} } S$", $\kappa$ is the
\emph{stoichiometry coefficient} and  the operator $\mbox{\texttt{op}}$  indicates the role of the species in the reaction $\alpha$. Specifically, $\mbox{ \texttt{op} }= \reactant$  denotes  a \emph{reactant},
$\product$ a \emph{product}, $\activator$ an \emph{activator},
$\inhibitor$ an \emph{inhibitor} and $\modifier$ a generic
\emph{modifier}, which indicates more complex roles than $\activator$ or $\inhibitor$.  The operator ``$+$'' expresses a choice
between possible actions.
Finally, the process $P \sync{\mathcal{L}} Q$
denotes the synchronisation between components: the set $\mathcal{L}$
determines those activities on which the operands are forced to
synchronise.  When $\mathcal{L}$ is the set of common actions, we use the shorthand notation $P \sync{*} Q$.

A Bio-PEPA system $\mathcal{P}$ is defined as a 6-tuple
$\langle \mathcal{V},\mathcal{N},\mathcal{K}, \mathcal{F}_R,Comp,P
\rangle$, where: $\mathcal{V}$ is the set of compartments,
$\mathcal{N}$ is the set of quantities describing each species (includes the initial concentration),
$\mathcal{K}$ is the set of all parameters referenced elsewhere, $\mathcal{F}_R$ is
the set of functional rates that define all required kinetic laws, $Comp$ is the set of
species components $S$ that highlight the reactions an entity can take part in and $P$ is the system model component.
Bio-PEPA models (i) represent reversible reactions as pairs of   irreversible forward and backward reactions, (ii) treat the same species in different states or compartments as different species represented by distinct Bio-PEPA components and (iii) assume static compartments. See \cite{CiocHill09}  for more details.

A variety of analysis techniques can be applied to a single Bio-PEPA model, facilitating the easy validation of analysis results when the
analyses address the same issues \cite{Calder062} and enhancing insight
when the analyses are complementary \cite{PASM08}. Currently supported
analysis techniques include stochastic simulation at the molecular
level, ordinary differential equations, probabilistic model checking
and numerical analysis of continuous time Markov chains
\cite{CiocHill09,BioPEPA,DesignAndDevelopment}.

\subsection{SBGN-PD mapping}
\label{sec: SBGN-PD mapping}
Here we map the core elements of SBGN-PD to Bio-PEPA (see \cite{SBGNtextWebsite} for an implementation).  

\textbf{Entity Pool Nodes}
Due to the rich encoding of information in the \texttt{EntityPoolNodeID}, Bio-PEPA can treat each distinct \texttt{EntityPoolNodeID} as a distinct species  component. This removes the need to explicitly consider any other aspects such as entity type, modifications, complex structures and compartments, as all such information is implicitly passed on to Bio-PEPA by using the \texttt{EntityPoolNodeID} as the name for the corresponding species component.  To define the set $\mathcal{N}$ of a Bio-PEPA system requires the attribute \verb"InitialMoleculeCount" for each EPN (see Section 3).

\textbf{Processes}
All SBGN-PD \verb"ProcessTypes" are simply represented as reactions in Bio-PEPA. Compiling the corresponding set $\mathcal{F}_R$ relies on the attribute \verb"PropensityFunction" and a substitution mechanism that makes it easy to define these functions manually. 
To help humans understand references to processes in the sets $\mathcal{F}_R$ and $Comp$ requires recognizable names for SBGN-PD \texttt{ProcessNodeID}s that map directly to their identifiers in Bio-PEPA. Thus graphical editors need to allow for manual \texttt{ProcessNodeID}s.

 {\em Reversible processes.} The translator supports reversible SBGN-PD processes by dividing them into two unidirectional processes for Bio-PEPA. The translator reuses the manually assigned \texttt{ProcessNodeID} and augments it by "\_F" for forward reactions and "\_B" for backward reactions.  These two unidirectional processes are then treated independently. 
 When compiling the species  components in Bio-PEPA, every time a   \verb"LeftHandSide" arc is found, the translator assumes that the corresponding forward and backward processes have been defined and will augment the process name by "\_F" for forward reactions and "\_B" for backward reactions, while adding the corresponding Bio-PEPA operator for reactant and product. \texttt{RightHandSide} arcs are handled in the same way. 
 Thus the production arc glyph in SBGN-PD has three distinct meanings as shown in Table 3.

\begin{small}
\begin{table}
\label{ArcsMapToBioPEPA}
\centering  
\extrarowheight4pt
\begin{tabular}{c  c  c  c c}
\hline
SBGN-PD glyph &  \; \; \verb"ArcType"  & \;\; Bio-PEPA symbol \;\; & Bio-PEPA code   \\[3pt]
\hline\\[-13pt]

 \begin{minipage}[c] {1.8cm} 
\includegraphics[width=1.0\textwidth] {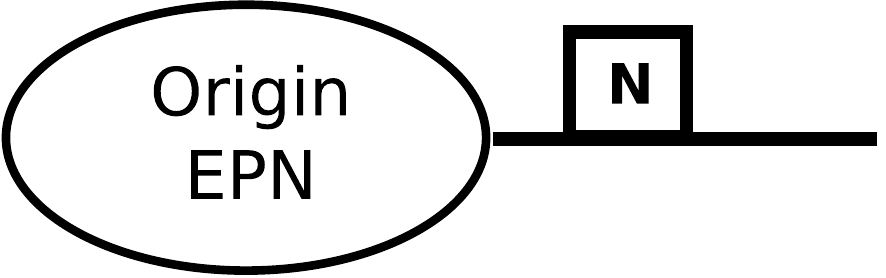}
\end{minipage}  &
 \verb"Consumption" & $\reactant$ & \verb"<<"  \\

 \begin{minipage}[c] {1.8cm} 
\includegraphics[width=1.0\textwidth] {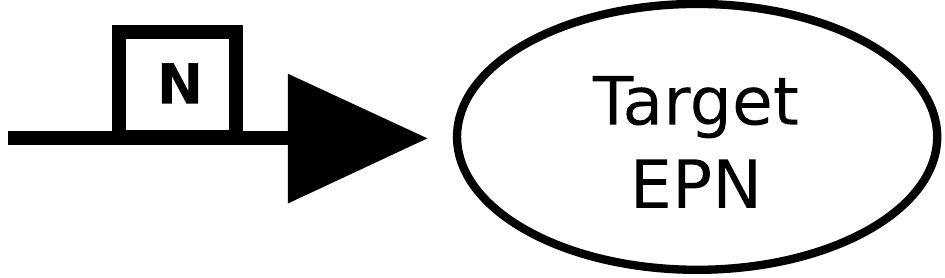}
\end{minipage}  &
 \verb"Production" & $\product$ & \verb">>"  \\

 \begin{minipage}[c] {1.8cm} 
\includegraphics[width=1.0\textwidth] {Table-Glyph-production.pdf}
\end{minipage}  &
\verb"LeftHandSide" &  $\reactant$ and $\product$  & \verb"<<" and \verb">>"   \\

 \begin{minipage}[c] {1.8cm} 
\includegraphics[width=1.0\textwidth] {Table-Glyph-production.pdf}
\end{minipage}  &
 \verb"RightHandSide" & $\product$ and $\reactant$  & \verb">>" and \verb"<<"   \\

 \begin{minipage}[c] {1.8cm} 
\includegraphics[width=1.0\textwidth] {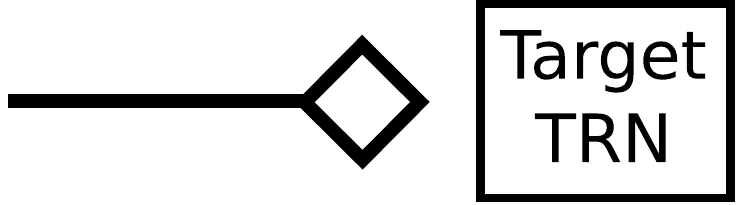}
\end{minipage}  &
\verb"Modulation" & $\modifier $ & \verb"(.)"  \\

 \begin{minipage}[c] {1.8cm} 
\includegraphics[width=1.0\textwidth] {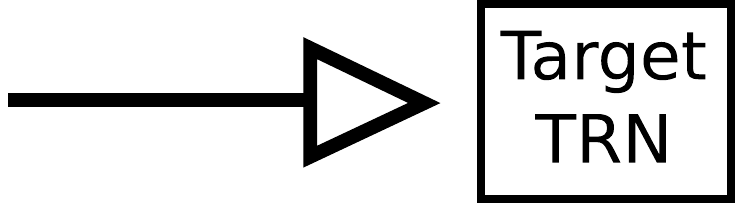}
\end{minipage}  &
\verb"Stimulation" & $\activator $ & \verb"(+)"  \\

 \begin{minipage}[c] {1.8cm} 
\includegraphics[width=1.0\textwidth] {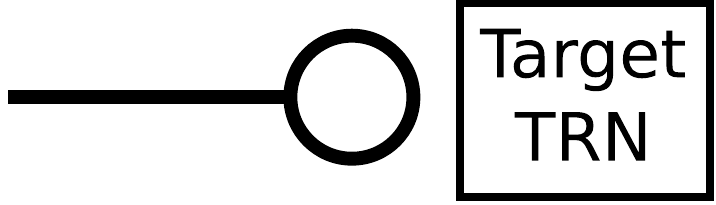}
\end{minipage}  &
\verb"Catalysis" & $\activator $ & \verb"(+)"  \\

 \begin{minipage}[c] {1.8cm} 
\includegraphics[width=1.0\textwidth] {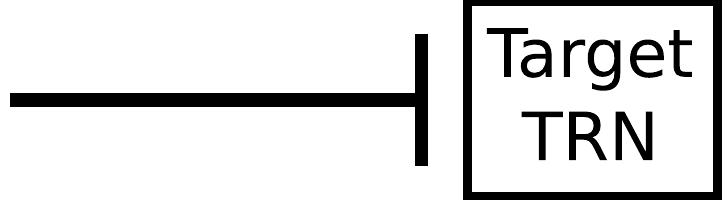}
\end{minipage}  &
\verb"Inhibition" & $\inhibitor $ & \verb"(-)"  \\

 \begin{minipage}[c] {1.8cm} 
\includegraphics[width=1.0\textwidth] {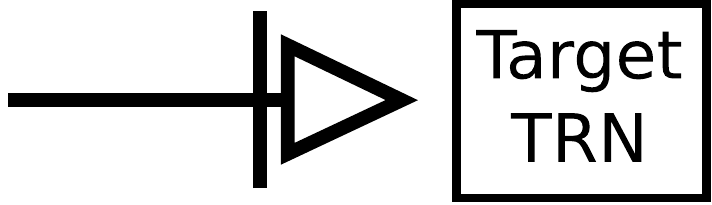}
\end{minipage}  &
\verb"NecessaryStimulation" & $\modifier $ & \verb"(.)"  \\[5pt]
\hline & &  \\[-12pt]

\end{tabular}
\caption{  ``Water pipes and control electronics'': Mapping arcs between entities and processes in SBGN-PD to operators in Bio-PEPA species  components. The ``symbols'' are the formal syntax of Bio-PEPA, while the ``code'' gives the concrete syntax used in the Bio-PEPA Eclipse Plug-in  \cite{BioPEPA}. } 
\end{table}
\end{small}

\textbf{Arcs.}
The arcs in SBGN-PD define which entities interact in which processes. Thus arcs play a pivotal role in defining the species  components in Bio-PEPA. Since arcs can store kinetic parameters, they are also important for defining parameters  in Bio-PEPA. 
As kinetic law definitions in Bio-PEPA frequently refer to such parameters, we use the ArcID that is automatically generated by the graphical editor to substitute the local manual arc references in propensity functions by globally unique parameters names (see Section 3). 
The type of an arc indicates both the role of the connected entity in the process (consumed reactant, product or modifier of rate) and the chemical nature of the reaction (catalysis, stimulation, inhibition, necessary stimulation or the most generic modification). 
Thus the type of an arc can be mapped directly to the operator ``$\mbox{\texttt{op}}$'' described in the Bio-PEPA syntax shown in Table~3.

\textbf{ Logical operators}
Logical operators require the conversion of integer molecule counts of the relevant EPNs to  binary signals and after some   boolean logic processing back to low and high integer values.  As evident from the implementation scheme above, the use of all quantitative properties culminates in the correct formulation of the corresponding propensity functions that determine the probability that the corresponding process will be the next to occur. Thus an implementation of logical operators requires that their results be included in the corresponding propensity functions. 
The current scheme of implementing propensity functions relies heavily on substituting the various components into the final equation, so that Bio-PEPA will ultimately only see one formula per propensity function. In this context the implementation of logical operators requires the insertion of a formula in the propensity function that computes the result of the boolean operations from their integer input. An arbitrarily complex logic operator network can be constructed from the following basic building blocks:

\begin{itemize}
\item \emph{Convert from integers or double floats to boolean values}. This is best done by a specially defined mathematical function that takes an integer or float signal and compares it to a specified threshold, giving back either 0 (signal $ <  $ threshold) or 1 (signal $ \ge $ threshold).  The definition of such a function is not complicated and is implemented in the current Bio-PEPA Eclipse Plug-in (starting with version 0.1).

\item AND operator. Multiply  all boolean inputs to arrive at the output.

\item NOT operator. The arithmetic expression (1- $input$) computes the  $output$

\item OR operator.  Sum all inputs (0/1) and test if it is greater than 1 using the threshold function. 

\end{itemize}

\noindent 
Of these, only the threshold function is not widely available, even if it is easy to implement.

\section{Converter implementation and internal representation}
\label{Sec:Converter}

We chose Java as implementation language for the converter described above, due to the good portability of the resulting binaries and the use of Java in the Bio-PEPA Eclipse Plug-in \cite{BioPEPA}.
 We defined a grammar for SBGNtext in the Extended Backus-Naur-Form (EBNF) as supported by ANTLR  \cite{ANTLR}, which automatically compiles the Java sources for the corresponding parser that stores all important parsing results in a number of coherently organised internal TreeMaps. To compile a working Bio-PEPA model three main loops over these TreeMaps are necessary: over all entities, over all processes and over all parameters. To illustrate the translation we refer to ``\texttt{code}'' examples from Figure 2.

The \textbf{loop over all entities} 
(e.g. ``\texttt{m\_MAPKK\_PP}'') compiles the species components  as well as the model description required by Bio-PEPA. The latter is a list of all participating \texttt{EntityPoolNodeID}s combined by the cooperation operator ``$<*>$" 
that automatically 
synchronises all common actions. This simplification depends on all processes in SBGN-PD having unique names and fixed lists of reactants  with no mutually exclusive alternatives in them. The first condition can be enforced by the tools that produce the code, the second is ensured by the reaction-style of describing processes in SBGN-PD.

For each species component a loop over all arcs finds the arcs that are connected to it (e.g. ``\texttt{st27}'') and that store all relevant \texttt{ProcessNodeID}s (e.g. ``\texttt{K\_P\_act}''). The same loop  determines the respective role of the component (as reflected by the choice of the Bio-PEPA operator in Table 3; e.g. ``\texttt{(+)}'').

The \textbf{loop over all processes} (e.g. ``\texttt{K\_P\_act}'') compiles the kinetic laws by substituting aliases (e.g. ``\texttt{<par: enz.kcat>}'' or ``\texttt{<ent: enz>}'')  for parameters (``\texttt{st27\_kcat}'') and EPNs (``\texttt{m\_MAPKK\_PP}'') in the propensity functions specified in the graphical editor. Each function is handled separately by a dedicated function parser that queries the TreeMaps generated when parsing the SBGNtext file.

The \textbf{loop over all quantitative properties}  of the model  defines the parameters in Bio-PEPA (e.g. ``\texttt{st27\_kcat}''). 
 It is possible to avoid this step by inserting the direct numerical values into the equations processed in the second loop. However, this substantially reduces the readability of equations in the Bio-PEPA code and makes it difficult for third party tools to assist in the automated generation of parameter combinations. Thus we defined a scheme that automatically generates parameter names to maximise the readability of equations (combine ArcID ``\texttt{st27}'' and name of the quantitative property ``\texttt{kcat}'').

To facilitate walking over the various collections specified above, the TreeMaps are organised in four sets, one for entities, one for processes, one for arcs and one for quantitative properties. Each of these sets is characterised by a common key to all maps within the set. This facilitates the retrieval of related parse products for the same key from a different map. Since all this information is accessible from Java code, it is easily conceivable to use the sources produced in this work for reading SBGNtext files in a wide variety of contexts. The system is easy to deploy, since it involves few files  and the highly portable ANTLR runtime library. Our converter is called SBGNtext2BioPEPA and sources are available \cite{SBGNtextWebsite}.

\begin{small}
\begin{figure}
\label{figSBGNmodel}
\begin{center}
\includegraphics[width=1.0\textwidth] {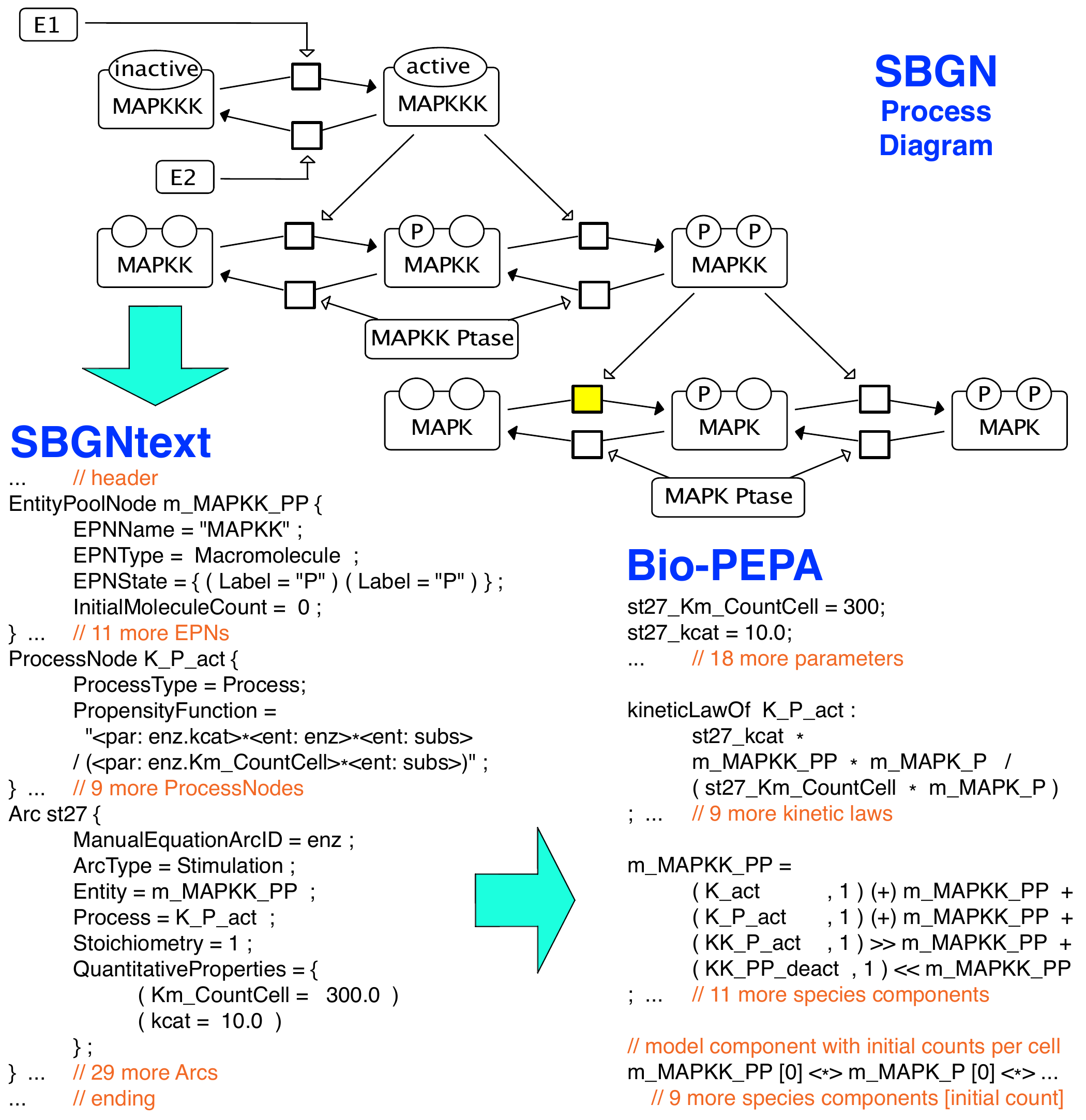}
\caption{A possible SBGN-PD representation of a MAPK cascade with fragments from the corresponding SBGNtext code and the automatically generated Bio-PEPA code. Most biologists would find it far more natural to draw the diagram at the top than to write the Bio-PEPA code. 
The code excerpts focus on EPN MAPK-PP and the reaction it catalyses (\texttt{'K\_P\_act'}, yellow node). All parameters are scaled to represent counts of molecules per cell (including the Michaelis-Menten constants). For the full code and newer versions see \cite{SBGNtextWebsite}.
Biologically, the input signal is a change in \texttt{E1}, the output signal a change in \texttt{MAPK\_PP}. The back reactions on each of the three levels can be thought of as a ``conveyer belt'' that constantly turns active  molecules (on the right) into passive ones (on the left). \texttt{E1} must overcome the conveyer belt on the first level for sequentially overcoming the other conveyer belts too. In many such systems either active or inactive molecules are essentially absent. } 
\end{center}
\end{figure}
\end{small}

\section{Example: Stochasticity in the MAPK cascade}
\label{Sec:Stochastiicity}

Here we illustrate our translation by applying SBGNtext2BioPEPA \cite{SBGNtextWebsite} to a real-world example. We implemented a simple model of the Mitogen-Activated Protein Kinase (MAPK) signal transduction cascade \cite{Kwiatkowska08,Huang96} in order to investigate the time needed for the cascade to switch from ``off'' to ``on''. The general pattern of the MAPK cascade is well conserved across many biological taxa and triggers a highly varied range of molecular responses \cite{Huang96}. Such a broad conservation suggests not only important functionality (maintained by purifying selection), but also an astonishing flexibility in how a MAPK cascade might be implemented (it has to work in a wide range of contexts). For example, MAPK cascades operate in oocytes of the frog \textit{Xenopus} and in yeast cells, despite their  5 million fold differences in size \cite{Huang96,Wallace78} and substantial differences in kinetic constants \cite{Huang96,Kholodenko2000}. 

Figure 2 shows a SBGN-PD diagram of the basic structure of a MAPK cascade. It accepts ``input'' in the form of a change in the concentration of the enzyme E1 and then propagates the signal by changing the concentrations of the various intermediate
 substrates and enzymes until the ``output'' enzyme MAPK\_PP is affected. If the input is 
off (low E1), the output is off too (no MAPK\_PP). If the input is on (E1 above threshold), output is reliably on as well (high MAPK\_PP). Important properties for signalling cascades include the reliability of transmitting a signal and the speed with which this happens. Building on the model in \cite{Huang96}, a recent study explored the expected percentage of active output and the time until all output is activated \cite{Kwiatkowska08}.

\begin{small}
\begin{figure}
\label{figModelE1eq20}
\begin{center}
\includegraphics[width=1.0\textwidth] {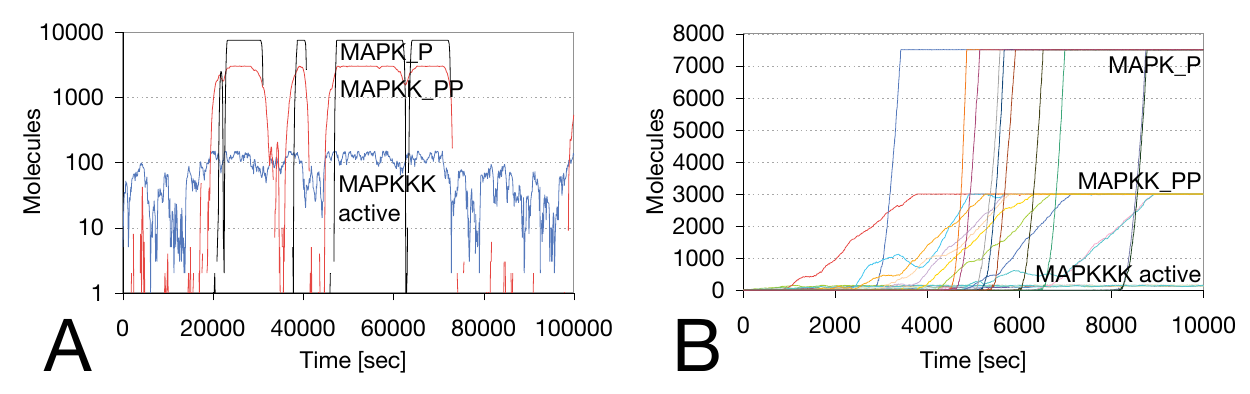}
\caption{Noise in the MAPK cascade. (A) One stochastic simulation trace, where E1 = E2 = 20 makes the cascade switch between the ``on'' and ``off''state, as activating and deactivating reactions at the first level are equally strong. 
(B) Ten stochastic simulation traces, where E1 = 21, E2 = 20. If activating reactions are only slightly stronger than deactivating reactions at the first level of the cascade, the cascade can be expected to switch ``on''. However the time until the ``on'' state is reached is subject to considerable stochasticity, which is mostly determined by stochasticity during the initial stages of the switching processes.} 
\end{center}
\end{figure}
\end{small}

\begin{small}
\begin{figure} [h!] 
\label{figModelE1eq40to100}
\begin{center}
\includegraphics[width=1.0\textwidth] {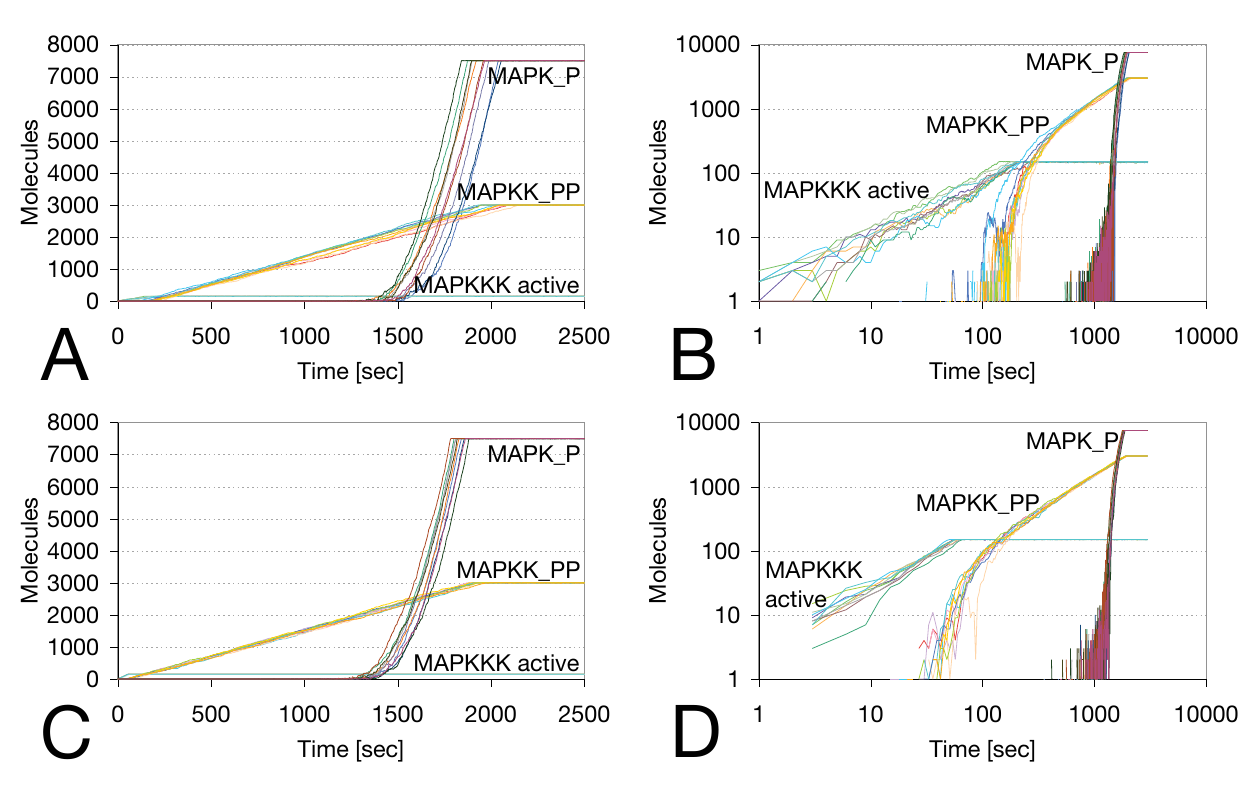}
\caption{Increasing the number of E1 molecules from 40 (upper part, A, B) to 100 (lower part, C, D) reduces the stochasticity of the time to the completion of the switch (see linear plots on the left, A, C). The stochasticity in the initial stages is considerable in both cases for this system (see log plots on the right, B, D). Ten stochastic simulation traces are shown. E2 = 20; for other parameters, see text.} 
\end{center}
\end{figure}
\end{small}

Here we automatically translate our SBGNtext model of the MAPK cascade via SBGNtext2BioPEPA into a Bio-PEPA model (see \cite{SBGNtextWebsite} for code), which we analyse with the Bio-PEPA Eclipse Plug-in   \cite{BioPEPA}. We report the results of stochastic simulations using the Gibson-Bruck algorithm \cite{Gibson00,Gillespie07}. Each reaction is assumed to follow Michaelis-Menten kinetics, resulting in the following propensity function:
$$\frac{k_{cat}[count/sec] * Enzyme[count] * Substrate[count]}{K_M[count] + Substrate[count]}$$

\noindent where [count] indicates the absolute number of molecules of this entity within the cell, $k_{cat}$ denotes the number of substrate molecules that can be processed by one enzyme at maximal speed, $K_M$ denotes the Michaelis-Menten constant that is an inverse measure of the affinity between enzyme and substrate. We assume $K_M$ = 300 and $k_{cat}$ = 10 for all reactions (see \cite{Huang96,Kholodenko2000} for other possible values). All our simulations start with molecule counts of E2 = 20, MAPKKK = 150, MAPKK = 3000, MAPKK\_Ptase = 100, MAPK = 7500,MAPK\_Ptase = 2000 and 0 for all other entities, reflecting the equilibrium ``off'' state. These molecule counts roughly reflect MAPK cascades in yeast, and are about 6 orders of magnitude below corresponding counts in \textit{Xenopus} oocytes \cite{Huang96}.  We tested various E1 input signals and found the following results  that cannot be obtained without translating SBGN-PD into a quantitative formalism. 

\begin{enumerate}
\item The cascade remains ``off'', if E1$<$E2  (not shown).
\item The cascade switches beween ``on'' and ``off'', if  E1 = E2 as shown in Figure 3A. 
\item The cascade switches reliably to ``on'' if E1$>$E2, but at E1 = 21 there is considerable noise in the ``signalling time'' (from switching E1 ``on'' until all output MAPK\_PP is switched ``on'' as well; see Figure 3B).
\item   Further increasing E1 substantially reduces the variability in signalling times and slightly improves speed (Figure 4).
\item Additional tests have shown that the overall signalling speed is very strongly influenced by $k_{cat}$ (not shown).

\end{enumerate}

\section{Related work}
\label{Sec:Related work}

There are various tools that map visual diagrams to quantitative modelling environments (e.g. SPiM \cite{Phillips09}, BlenX \cite{BlenX}, kappa \cite{Danos07}, Snoopy \cite{Snoopy}, EPN-PEPA \cite{EPN2BioPEPA}, JDesigner \cite{JDesigner} ). However the corresponding graphical notations are not as rich as SBGN-PD and are thus not easily applied to the wide range of scenarios that SBGN-PD was designed for. Since SBGN-PD is emerging as a new standard, it is clearly desirable to translate from SBGN-PD to a quantitative environment.  

	Since the first draft of SBGN-PD has been published in August 2008, a number of  tools are being developed to support it.
	The graphical editor CellDesigner \cite{CellDesigner} supports a subset of SBGN-PD and can translate it into SBML which is supported by many quantitative analysis tools. However the process of adding quantitative information involves cumbersome manual interventions. This motivated work for SBMLsqueezer \cite{SBMLsqueezer}, a CellDesigner plug-in that supports the automatic construction of generalised mass action kinetics equations. While the automated suggestions for the kinetic laws from SBMLsqueezer might be of interest for some problems, the generated reactions contain many parameters that are extraordinarily difficult to estimate. Thus it is preferable to also allow the user to enter arbitrary kinetic laws that may have to be hand-crafted, but whose equations are simpler and require fewer parameter estimates. In SBGNtext2BioPEPA this is combined with mechanisms to reuse the code for such kinetic laws,  greatly reducing practical difficulties and the potential for errors.

\section{Conclusion and Perspectives}
\label{Conclusion}
We have explicitly described the process flow abstraction that implicitly underlies SBGN-PD. We have used this abstraction to design  a mechanism for translating SBGN-PD into code that can be used for quantitative analysis, using Bio-PEPA as an example. In order to do this we build on SBGNtext, a textual representation of SBGN-PD that we created  \cite{SBGNtextWebsite,SBGNtextTechReport} and that focusses on the key functional SBGN-PD content, avoiding the clutter that comes from storing graphical details. 
We have developed our translator SBGNtext2BioPEPA in Java to facilitate its integration in the Systems Biology Software Infrastructure that is currently under development at the Centre for Systems Biology at Edinburgh (http://csbe.bio.ed.ac.uk/). SBGNtext2BioPEPA contains a parser for our SBGNtext format based on a formal ANTLR EBNF grammar and is freely available  \cite{SBGNtextWebsite}. Building on the process flow abstraction and the internal representation of entities, processes, arcs and parameters in our code can make it easy to implement translations of SBGNtext to other modelling platforms.  Since biologists are much more comfortable with drawing visual diagrams than writing code, support for translating SBGN-PD into quantitative analysis systems can play a key role in facilitating quantitative modelling.

\label{sec:Acknowledgement}
\paragraph{Acknowledgements.} 
We thank Stephen Gilmore for comments that improved this manuscript. The Centre for Systems Biology Edinburgh is a Centre for Integrative
Systems Biology (CISB) funded by BBSRC and EPSRC, reference BB/D019621/1.

\bibliographystyle{eptcs} 

\end{document}